\documentclass[twocolumn,twocolappendix,astrosymb,tighten,apjl]{openjournal}
\usepackage{import}
\usepackage{tikz}
\usepackage{colortbl}
\usepackage{array}
\usepackage[newfloat]{minted}
\usepackage{listings}
\usepackage{pythonhighlight}
\usepackage{hyperref}
\usepackage{etex}

\usepackage{amssymb}
\usepackage{pifont}
\newcommand{\cmark}{\ding{51}}
\newcommand{\xmark}{\ding{55}}

\usepackage{graphicx}

\usepackage{subcaption}
\makeatletter
\expandafter\let\csname longtable*\endcsname\relax
\expandafter\let\csname endlongtable*\endcsname\relax
\makeatother

\usepackage{tabularx}
\usepackage{booktabs}
\usepackage{enumitem}
\usepackage{orcidlink}

\usetikzlibrary{positioning}
\usetikzlibrary{calc}
\usetikzlibrary{shapes}
\usetikzlibrary{arrows.meta}
\usetikzlibrary{bending}
\begin{document}

\title{OpenCosmo: Community Portal and Analysis Framework for Flagship Cosmological Simulations}

\author{
    Patrick R. Wells$^{1}$\orcidlink{0000-0003-0999-2395},
    Michael Buehlmann$^{2,1}$\orcidlink{0000-0002-8469-4534},
    Patricia Larsen$^{2}$\orcidlink{0000-0001-9592-4676},
    William M. Hicks$^{1}$\orcidlink{0009-0002-7068-6602},
    Manpreet Dhillon$^{1}$\orcidlink{0009-0005-6245-4547},
    Idunnuoluwa A. Adeniji$^{3,4}$\orcidlink{0009-0008-8644-103X},
    Katrin Heitmann$^{1}$\orcidlink{0000-0003-1468-8232},
    Salman Habib$^{2,1}$\orcidlink{0000-0002-7832-0771},
    Benoit C\^ot\'e$^{4}$\orcidlink{0000-0002-9986-8816},
    Thomas Uram$^{4}$\orcidlink{0000-0002-5631-0142},
    Gideon McFarland$^{5,6,1}$\orcidlink{0000-0003-2541-9433},
    Andrew Hearin$^{1}$\orcidlink{0000-0003-2219-6852},
    Ezar Shinbaro$^{1}$\orcidlink{0009-0009-5868-4322},
    Michael E. Papka$^{4,3}$\orcidlink{0000-0002-6418-5767}
}

\affiliation{$^{1}$High Energy Physics Division, Argonne National Laboratory, Lemont, IL 60439}
\affiliation{$^{2}$Computational Science Division, Argonne National Laboratory, Lemont, IL 60439}
\affiliation{$^{3}$Department of Computer Science, University of Illinois Chicago, Chicago, IL 60607}
\affiliation{$^{4}$Argonne Leadership Computing Facility, Argonne National Laboratory, Lemont, IL 60439}
\affiliation{$^{5}$Department of Physics and Astronomy, Northwestern University, Evanston, IL 60208}
\affiliation{$^{6}$Center for Interdisciplinary Exploration and Research in Astrophysics (CIERA), Northwestern University, Evanston, IL 60201}


\begin{abstract}
Cosmology is a precision observational science, and large simulations are necessary components of many analyses. These simulations are computationally expensive and produce massive, complex datasets; sharing them widely -- to enable further explorations, comparison with observations, and communication with general audiences -- is crucial to realizing their scientific value. In this paper, we introduce the OpenCosmo project, which provides flexible access to, and analysis of, flagship cosmological simulations performed with HACC. A web-based portal (\url{https://opencosmo.science}) serves custom subsets -- halo catalogs, profiles, particles, galaxy catalogs, and lightcone catalogs and maps -- from simulations including the two-trillion-particle Frontier-E gravity-only run, Last Journey, Discovery, and a 64-member hydrodynamic suite. A companion Python toolkit analyzes the returned data and scales without modification from laptop-sized subsets to full simulations on supercomputers. OpenCosmo supports multiple levels of interaction, from browser-based search and download to programmatic and AI-agent-driven workflows, by integrating with existing high-performance computing and data infrastructure. Its architecture, built on Globus services, provides a scalable and adaptable framework that can be extended to other scientific domains seeking to couple data sharing with computational capability.
\end{abstract}



\section{Introduction}

Extracting new scientific information from large datasets is one of the key challenges facing astrophysics and cosmology in the modern era. In this context, data access is often the foundational problem on which all else rests. But even if datasets are accessible, it is often challenging to analyze them further at scale. Addressing this widely recognized problem in an effective way increases both scientific productivity and the quality and reproducibility of the analyses performed. In this paper we present the OpenCosmo project, a multi-facility data analysis platform for cosmology that allows scientific users to request data from and carry out analyses on extreme-scale datasets without requiring detailed computational expertise. The first release provides structured access to large datasets from simulations performed with the Hardware/Hybrid Accelerated Cosmology Code (HACC) \citep{habib_2016} and its hydrodynamic extension CRK-HACC \citep{frontiere_2023} as well as associated downstream data products.

The OpenCosmo portal is a web interface for querying data from flagship cosmological simulations. Requests are automatically delegated to the facility housing the data, and results are returned as an HDF5 file with a custom structure. We provide a Python library built on \texttt{h5py}\footnote{\url{https://h5py.org}} that leverages this structure to provide data management and analysis capabilities to our users.

OpenCosmo has been released to early users within the Cosmological Physics and Advanced Computing (CPAC) group at Argonne National Laboratory. Topics studied include early dark energy, galaxy cluster dynamics, the effects of subgrid parameters on hydrodynamic simulations, and agentic frameworks for astrophysical and cosmological analyses. The data querying software has also been used within the LSST Dark Energy Science Collaboration\footnote{\url{https://lsstdesc.org/}} (DESC) to analyze synthetic galaxy catalogs produced in the \texttt{diffsky}\footnote{\url{https://github.com/ArgonneCPAC/diffsky}} project. The software has been readily adopted by students; because the toolkit hides the details of distributed computing, a script developed as a single process on a small dataset can be moved to a large high-performance computing (HPC) system and run across a flagship simulation with no changes. This is advantageous for early-career and established researchers alike, who can now focus their time on expressing their scientific ideas and interpreting results. 

The OpenCosmo portal and data tooling are now ready for general use, and the portal is available at \url{https://opencosmo.science}, subject to certain access requirements (see Section~\ref{sec:access}). A website with curated example notebooks is available at \url{https://argonnecpac.github.io/opencosmo-examples/}. In this paper, we discuss the tools that are currently available, the system architecture, and accessible datasets. The datasets include some that have been previously released, as well as several new simulations exploring variations in subgrid physics. 
The present iteration supports a set of specific, predefined queries, which are accessible through multiple interfaces: the web portal, a command-line client, and Model Context Protocol (MCP) servers that expose the platform's capabilities as tools for AI agents (Section~\ref{sec:programmatic}). Significantly more powerful access modes are under active development, including remote access to full-scale datasets through a Python API, currently in closed beta testing.
We are exploring frameworks that will allow users to run full-scale analysis pipelines on petabyte-scale datasets without having to manage software environments on an HPC system, write job submission scripts, or even engage directly with the system (e.g., via SSH). These capabilities are increasingly being built on top of infrastructure from the American Science Cloud \citep{doe_asc_2025}, which accelerates multi-facility application development.

We begin with a discussion of the project's goals, comparisons to existing frameworks, an overview of its major components, and a full example workflow in Section~\ref{sec:overview}. In the sections that follow, we discuss the individual components of the platform, starting with the simulations and data products available to users via the portal in Section~\ref{sec:sims}. We describe the OpenCosmo web portal in Section~\ref{sec:tools}, including its design philosophy and user experience, the available query tasks, its interactive visualization capabilities, its programmatic interfaces, and access requirements. Section~\ref{sec:technical} focuses on the implementation, including data formats and architecture, and we conclude in Section~\ref{sec:conclusion} with further discussion and ideas for future evolution. 

\section{Project Goals and Overview}
\label{sec:overview}

The OpenCosmo project addresses a fundamental problem in data-driven science: as datasets grow in size and complexity, so does the difficulty of extracting science from them. Working with petabyte-scale data requires expertise in data management frameworks and distributed computing, as well as access to significant compute and storage resources -- skills and resources largely disjoint from those needed for scientific analysis.

One response to this gap is education: train individual scientists in the tools and techniques for working with large datasets on HPC systems. While some level of training is essential, we take a complementary approach. OpenCosmo provides a platform for data access and analysis orchestration that implements best-practice data management and orchestration routines once, on behalf of all its users, so that scientific analyses can be expressed independently of data scale and location.

An additional driver is generalizability;  we seek to develop reusable frameworks that can also be used to build data platforms for other science domains. In our view, cross-facility, managed science workflows (or ``Science as a Service'') are an essential tool for accelerating progress. These types of platforms must understand the needs of the specific domain they serve, but many of the individual infrastructure pieces can be generalized and used much more widely.

\subsection{Goals and Comparison to Existing Frameworks}

Data access is a core problem in any science, and we are by no means the first team to seek solutions for cosmological data. For example, the NSF NOIRLab Astro Data Lab~\citep{noir_datalab} provides access to a wide variety of large survey datasets as well as some simulated datasets. The Rubin Science Platform~\citep{rsp} serves data specifically from the Vera C. Rubin Observatory, and provides compute in the form of cloud-based Jupyter notebooks among other tools. CosmoHub~\citep{cosmohub} is similar to the Astro Data Lab in that it provides direct access to a large variety of datasets, but has paid particular attention to an expressive user interface.

These platforms primarily serve data in response to queries, through interfaces that are generally wrappers around SQL or SQL-derived languages such as the Astronomical Data Query Language. The user decides what to download, and defines analysis tasks on their own resources with their own software. This model breaks down when an analysis requires substantial compute or more data than can realistically be downloaded: the analysis framework must then load and unload data during the analysis itself and coordinate compute across distributed resources. In this regime, the platform must bring the analysis to the data, rather than the data to the analysis.

Platforms for managed scientific workflow execution also exist. Within astrophysics, a prominent example is the NASA Fornax Initiative\footnote{\url{https://science.nasa.gov/astrophysics/programs/physics-of-the-cosmos/community/the-fornax-initiative/}}, which aims to provide managed cloud compute with direct access to NASA data.  Other examples from the broader scientific community include nanoHUB~\citep{klimeck2008_nanohub}\footnote{\url{https://nanohub.org/}} for computational nanotechnology research and the Galaxy Project~\citep{galaxy_project} for general scientific workflow management. These platforms enable users without deep technical backgrounds to perform complex analyses in managed compute environments, but the available compute is typically limited and oriented toward interactive, small-scale workflows, and large datasets are usually exposed only through SQL queries that are difficult to integrate into an analysis pipeline.

The OpenCosmo project combines these two capabilities: a fully managed platform for analyzing petabyte-scale cosmological datasets on DOE leadership-class computing facilities, designed to support both large-scale analysis workloads and users who have never run a multi-node workflow. The same tooling runs unchanged from laptops to leadership-class systems. Sensible default parameters and a small number of required decisions give users unfamiliar with cosmological data quick access to relevant subsets, while the underlying interfaces allow advanced users to express complex analysis tasks.

The OpenCosmo portal represents the first user-facing service released as part of this platform. It provides easy, web-based access to flagship cosmological simulation data and a set of query tasks designed to retrieve subsets of data that are relevant to a user's science. We are in the process of building advanced capabilities, such as a Python API for expressing arbitrary queries remotely. This portal and the larger platform are built on top of several distinct but interlocking pieces. 

\subsection{Primary Components}

The first OpenCosmo release is built around four primary components, which together provide a unified platform for data access and high-quality tooling for performing complex analysis workflows on the available data. We introduce each component briefly here and discuss them in detail in later sections.

\paragraph{Data} The core of the OpenCosmo platform is data from state-of-the-art cosmological simulations performed with HACC and CRK-HACC, together with derived data products. The size of these datasets is the principal obstacle to their scientific use, and serving them in a form that can be worked with at scale is the central goal of the project. The available simulations and data products are described in Section~\ref{sec:sims}.

\paragraph{The OpenCosmo Toolkit} The OpenCosmo toolkit is an open-source\footnote{\url{https://github.com/ArgonneCPAC/OpenCosmo}} Python library that provides programmatic access to simulation data (and analysis products) produced by the CPAC team, with expansion to other datasets planned in the near future. Although originally conceived to help users work with data returned to them by the web portal, it handles the underlying full-scale simulations and distributed analysis workflows; all queries available on the portal are performed by the toolkit. It supports automatic linking of related data types, allowing users to address complex multi-modal datasets with a single, unified interface (Section~\ref{sec:toolkit}).

\paragraph{Web Portal} The OpenCosmo portal (\url{https://opencosmo.science}; see Section~\ref{sec:access} for access requirements) is the web interface for browsing datasets and submitting queries. It provides a set of standard, parameterized query tasks covering a wide variety of use cases, with results typically returned within minutes, enabling near real-time interaction (Section~\ref{sec:tools}).

\paragraph{Orchestration Layer} Once a query is submitted on the web portal, it is handed off to a multi-facility orchestration layer built on the Globus Compute platform. This layer examines the query to determine which facility it should be delegated to, submits jobs to the appropriate HPC system, monitors their status, and communicates with the user once they complete. Datasets are available for download immediately, or can be transferred via Globus Transfer~\citep{chard2014_globus} to the user's preferred resources. The orchestration layer is entirely agnostic to the actual computational workload being performed, and does not interact with the data directly; it is described in more detail in Section~\ref{sec:arch}.

\subsection{Example Workflow}
\begin{figure*}[t]
\centering

\newlength{\topindent}
\setlength{\topindent}{0.08\textwidth} 

\begin{minipage}[t]{0.52\textwidth}
    \vspace{0pt}
    \centering
    \includegraphics[width=\linewidth]{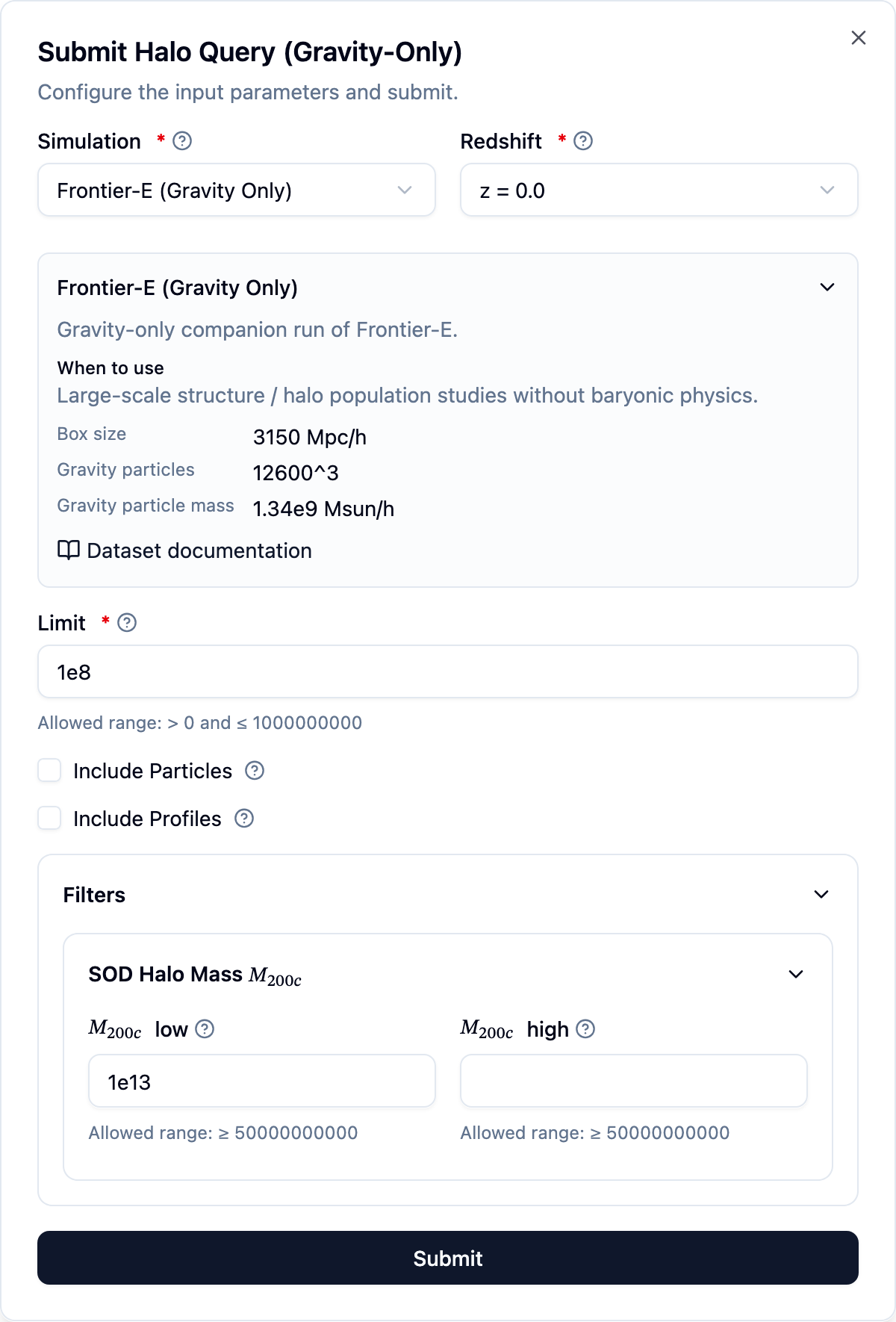}
\end{minipage}\hfill
\begin{minipage}[t]{0.46\textwidth}
    \vspace{0pt}

    \begin{minted}{python}
import opencosmo as oc
import numpy as np
import matplotlib.pyplot as plt

ds = oc.open("filtered_haloproperties.hdf5")
mass = ds.select("sod_halo_mass").get_data("numpy")
volume = ds.simulation["box_size"]**3

bins = np.logspace(13, 15.5)
log_bins = np.log10(bins)
bin_centers = np.sqrt(bins[1:] * bins[:-1])
dlogm = np.diff(log_bins)

mass_hist, _ = np.histogram(mass, bins)
hmf = mass_hist / dlogm / volume

fig, ax = plt.subplots()
ax.plot(bin_centers, hmf)
ax.set(xscale="log", yscale="log")
    \end{minted}

    \vspace{0.75em}

    \centering
    \includegraphics[width=\linewidth]{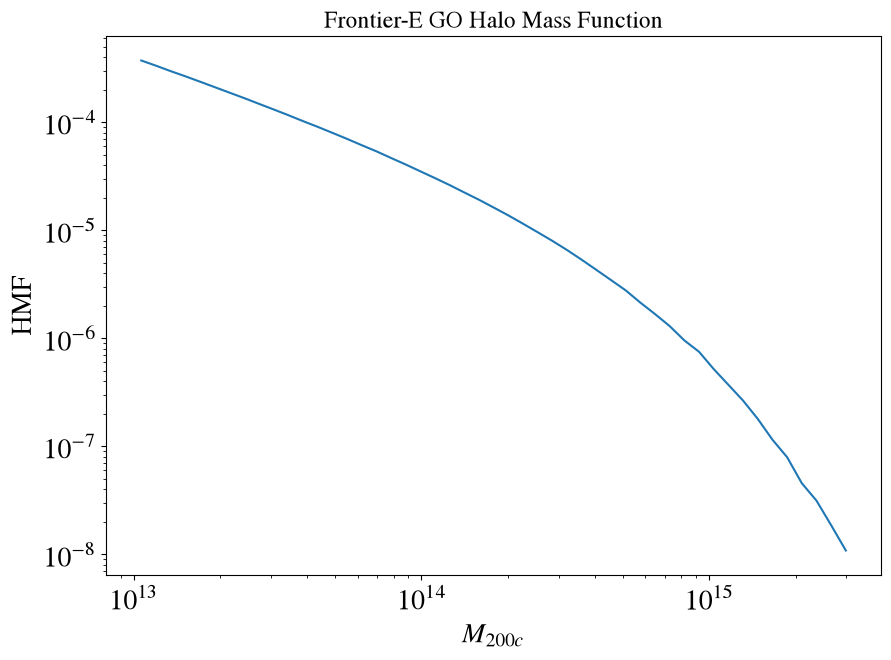}
\end{minipage}
\caption{End-to-end workflow for computing the halo mass function for all halos in the Frontier-E GO (gravity-only) simulation with mass greater than $10^{13}\,\mathrm{M}_\odot$.
Left: The web portal query interface, where the user selects the
Frontier-E GO simulation, redshift $z=0$, and a group-scale halo mass constraint ($M_{200c} \geq 10^{13}\,\mathrm{M}_\odot$), while limiting the results to a maximum of $10^8$ halos.
Right: Code snippet demonstrating data retrieval and the computation and visualization of the halo mass function.}
\label{fig:workflow}

\end{figure*}

Figure~\ref{fig:workflow} shows a complete example workflow. The user submits a halo catalog query to the web portal, selecting halos with masses above $10^{13}\,\mathrm{M}_\odot$. The query completes in approximately 5 minutes and produces a 2.6\,GB HDF5 file that is immediately available to download. Loading the file with the OpenCosmo toolkit and retrieving the halo masses and simulation box volume requires two lines of code; the halo mass function is then computed with \texttt{numpy}~\citep{numpy}, and plotted with \texttt{matplotlib}~\citep{matplotlib} (code setting figure and axis titles is omitted for readability). The active work of submitting the query and producing the plot takes less than 10 minutes, with roughly 10 minutes added for the query to complete and the results to download. The query returned $\sim 1.7\times10^7$ halos, well below the $10^8$-halo limit set in the query, indicating that the sample is complete for this mass range. Although the file is fairly large, only the halo-mass column is loaded into memory on the user's machine, keeping the memory footprint minimal. A natural continuation of this analysis would be to identify extreme objects from other catalog properties and retrieve their particle data for deeper study. 

\section{Simulations and Data Products}
\label{sec:sims}

Simulation datasets available on the portal have been generated with HACC~\citep{habib_2016}, a high-performance, GPU-accelerated, gravity-only cosmology code designed for survey-scale simulations carried out on leadership-class HPC platforms, and CRK-HACC~\citep{frontiere_2023}, a major extension that adds gas dynamics, cooling and heating mechanisms, models for black hole and galaxy formation, and includes a number of astrophysical feedback mechanisms such as supernova and active galactic nuclei (AGN) feedback~\citep{frontiere2025_galaxyformation}.

The initial OpenCosmo data tranche includes results from four large-scale gravity-only simulations with both $\Lambda$CDM and $w_0$--$w_a$ cosmologies, a synthetic galaxy catalog based on modeling the galaxy--halo connection on top of gravity-only simulation results~\citep[see, e.g.][]{diffmah, diffstar}, and 64 smaller-scale hydrodynamic simulations. We plan to release further simulation-based datasets in the future, including the extreme-scale Frontier-E hydrodynamic simulation described in~\citet{2025arXiv251003557F}. We discuss the available simulations in more detail in the following sections. A summary of available simulations and data products can be found in Table~\ref{tab:param}. We first describe the available simulations and catalogs, and then turn to the data products derived from them.

\begin{table*}[t]
    \caption{Summary of available simulations and data products.  $L_\mathrm{box}$ is the comoving box side length and $m_p$ the particle mass; for the SciDAC 128 SG5 hydrodynamic suite, the two masses are the CDM and initial baryon particle masses. Diffsky is a synthetic galaxy catalog built on Last Journey and therefore lists no box size or particle mass; the Discovery row covers both the $\Lambda$CDM and $w_0$--$w_a$ runs. A check mark indicates that the corresponding data product can currently be queried through the portal; the ``Sec.'' column points to the subsection describing each dataset.}
    \begin{tabular*}{\textwidth}{@{\extracolsep{\fill}}lccccccc@{}}
    \toprule
         Simulation(s)/Catalog & Sec. & $L_\mathrm{box}$ & $m_p$ & \multicolumn{4}{c}{Data products} \\
         \cmidrule(lr){5-8}
         & & [Mpc] & [$\mathrm{M}_\odot$] & Halo Properties & Particles/Profiles & Galaxies & Maps \\
    \midrule
    
        Frontier-E GO & \ref{sec:frontier-e} & 4,660 & $2\times 10^9$ & \cmark & \cmark & \xmark & \cmark \\
        
        Last Journey & \ref{sec:lastjourney} & 5,025 & $4\times 10^9$ & \cmark & \cmark & \xmark & \xmark \\

        Diffsky & \ref{sec:diffsky} & N/A & N/A & \xmark & \xmark & \cmark\tablenotemark{$\dagger$} & \xmark \\

        Discovery & \ref{sec:discovery} & 1,500 & $4.4\times 10^8$ & \cmark & \xmark & \xmark & \xmark  \\

        SciDAC 128 SG5 Suite & \ref{sec:hydro} & 189.2 & $1.7\times 10^9$ / $3.2\times 10^8$ & \cmark & \cmark & \cmark & \xmark \\
        
        SciDAC 128 GO & \ref{sec:hydro} & 189.2 & $2\times 10^9$ & \cmark & \cmark & \xmark & \xmark \\
    \bottomrule
    \end{tabular*}
    
    \raggedright\footnotesize
    \vspace{0.5ex}
    $^{\dagger}$Synthetic galaxies, provided through the separate Diffsky catalog (see Section~\ref{sec:diffsky}).\\
    
    \label{tab:param}
\end{table*}

Unless otherwise noted, lengths are comoving, and we quote box sizes and particle masses without factors of $h$ (i.e., in Mpc and $\mathrm{M}_\odot$), converted from HACC's native comoving $h^{-1}\,\mathrm{Mpc}$ and $h^{-1}\,\mathrm{M}_\odot$ units using each simulation's Hubble parameter. The data products themselves store quantities in HACC's native units; the OpenCosmo toolkit tracks units explicitly, presents data in $h$-free comoving units by default, and converts to other conventions on request.

\subsection{Frontier-E Gravity-Only Simulation}
\label{sec:frontier-e}

HACC and CRK-HACC were recently used to carry out the Frontier-E simulations~\citep{2025arXiv251003557F} on the GPU-accelerated Frontier system at the Oak Ridge Leadership Computing Facility (OLCF). This pair of simulations consists of (1) the largest hydrodynamic simulation carried out to date, evolving just over four trillion particles and (2) an accompanying matched gravity-only simulation, evolving $12{,}600^3$ particles in a (4.66\,Gpc)$^3$ simulation volume. The cosmological parameters are based on the best-fit Planck cosmology~\citep{planck_2018}, \{$\Omega_c$, $\Omega_b$, $\sigma_8$, $n_s$, $h$\} = \{$0.26067$, $0.04897$, $0.8102$, $0.9665$, $0.6766$\}, leading to a (dark matter) mass resolution of $m_p\sim 2\times 10^9\,\mathrm{M}_\odot$. As part of our first OpenCosmo release, we provide access to the Frontier-E gravity-only simulation. A future release will also include results from the hydrodynamic simulation.

\subsection{Last Journey Gravity-Only Simulation}
\label{sec:lastjourney}

Last Journey~\citep{2021ApJS..252...19H} is a large-scale gravity-only simulation, evolving $10{,}752^3$ particles in a $(5.025\,\mathrm{Gpc})^3$ box in a Planck-like cosmology~\citep{planck_2018}, resulting in a mass resolution of $m_p \sim 4\times 10^9\,\mathrm{M}_{\odot}$. This simulation is targeted toward current cosmological surveys that simultaneously require resolution high enough to capture galaxies of $\sim 0.1L^*$ luminosities and volumes large enough to resolve rare cluster-mass objects. Of particular note, this simulation underlies the Diffsky dataset detailed in Section~\ref{sec:diffsky}. 

A selection of Last Journey data products was made available publicly in 2021 using the HACC data portal~\citep{hacc_data_portal}. Alongside the newly released simulations we provide a wider range of Last Journey data products through the OpenCosmo portal, allowing easier community access. Nine snapshots spanning $0 \leq z \leq 1.5$ are available ($z = 0$, $0.05$, $0.21$, $0.50$, $0.54$, $0.78$, $0.86$, $1.43$, and $1.49$).

\subsection{Diffsky Synthetic Galaxy Catalog}
\label{sec:diffsky}

The Last Journey simulation is complemented by a synthetic galaxy catalog based on models of the galaxy--halo connection. The synthetic galaxies are produced by Diffsky, a model that connects the mass assembly history (MAH) of simulated halos to the spectral energy distribution (SED) of galaxies that co-evolve with the halos. An early prototype of Diffsky has previously been used as part of the OpenUniverse project~\citep{openuniverse2024} to create synthetic catalogs based on the Outer Rim $N$-body simulation~\citep{heitmann_etal19_outer_rim}; new synthetic catalogs based on an updated formulation of the model have recently been used to graft synthetic galaxy SEDs onto the MAHs of halos in the Last Journey simulation~\citep{2021ApJS..252...19H}.

The particular $N$-body data products used by Diffsky to create synthetic galaxy catalogs are the core merger-trees, which are a computationally efficient alternative to conventional subhalo tracking algorithms~\citep{2023OJAp....6E..24K, sultan_etal21_last_journey2}. Diffsky was designed from the ground up to natively work with the core merger-trees, and so the properties of each synthetic galaxy in the Diffsky catalogs reflect the full assembly history of its associated core. Diffsky data products include precomputed photometry in a range of bands relevant to contemporary galaxy surveys such as the Vera C. Rubin Observatory Legacy Survey of Space and Time~\citep[LSST;][]{Ivezic_etal19_lsst_science}\footnote{\url{https://rubinobservatory.org}} and the Nancy Grace Roman Space Telescope\footnote{\url{https://roman.gsfc.nasa.gov}}; we additionally include sufficient information to recompute the high-resolution SED and derive photometric flux in any band. OpenCosmo-formatted Diffsky data are already being used regularly within DESC.

\subsection{Discovery Gravity-Only Simulations}
\label{sec:discovery}

The Discovery Simulations~\citep{2025OJAp....8E..74B} are a suite of gravity-only simulations aiming to probe the effect of evolving dark energy on structure formation, and directly inspired by recent results from the Dark Energy Spectroscopic Instrument (DESI)~\citep{2025JCAP...02..021A} and the Dark Energy Survey~\citep{2025arXiv250306712D}. Simulation data products are already available in full on the HACC data portal~\citep{hacc_data_portal}, and we also publish two of these simulations on the OpenCosmo portal. The first simulation is based on the current best-fit $\Lambda$CDM cosmology and the second includes the effects of a dynamical dark energy equation of state, parameterized via $w_0$ and $w_a$. Each simulation evolves $6{,}720^3$ particles in a (1.5\,Gpc)$^3$ box, leading to a mass resolution of $m_p\sim 4.4\times 10^8\,\mathrm{M}_\odot$. More specifically, the cosmology for the first simulation is given by \{$\Omega_c$, $\Omega_b$, $\sigma_8$, $n_s$, $h$, $w_0$, $w_a$\} = \{$0.2578$, $0.049$, $0.8135$, $0.968$, $0.6767$, $-1$, $0$\} and for the second simulation by \{$\Omega_c$, $\Omega_b$, $\sigma_8$, $n_s$, $h$, $w_0$, $w_a$\} = \{$0.2903$, $0.054$, $0.7917$, $0.9650$, $0.6466$, $-0.45$, $-1.79$\}.

\subsection{Hydrodynamic Simulations}
\label{sec:hydro}
Compared to their gravity-only counterparts, hydrodynamic simulations are much richer in the included baryonic physics and treatment of astrophysical processes. However, this comes at the cost of implementing many of these mechanisms in terms of parameterized phenomenological subgrid models -- the detailed physics cannot be modeled from first principles given the unavoidable spatio-temporal resolution limitations of the simulations. 

To enable the systematic investigation of astrophysical effects as a function of the  subgrid model parameters of CRK-HACC, we release results from 64 hydrodynamic simulations, collectively the SciDAC 128 SG5 suite (a $(128\,h^{-1}\mathrm{Mpc})^3$ box with five varied subgrid parameters). Each simulation has the same realization of the initial conditions to facilitate comparisons across the ensemble, and the cosmological parameters are the same as for the Frontier-E simulation, described in Section~\ref{sec:frontier-e}. Each simulation has a box volume of (189.2\,Mpc)$^3$ and evolves $2\times 512^3$ particles, leading to a CDM tracer particle mass of  $m_c\sim 1.7\times 10^9\,\mathrm{M}_\odot$ and a baryon mass resolution of $m_b\sim 3.2\times 10^8\,\mathrm{M}_\odot$. Given the relatively small volume of these simulations, we do not generate lightcone outputs.

The suite is accompanied by a single matched gravity-only simulation, evolving $512^3$ particles from the same initial conditions with a particle mass of $m_p \sim 2\times 10^9\,\mathrm{M}_\odot$. Because all 64 hydrodynamic realizations share the same initial conditions, this companion serves as a baseline for the entire suite, allowing the effects of baryonic subgrid physics to be separated from gravity-driven structure formation. Halo catalogs, profiles, and halo particles are available for the companion at the same snapshot redshifts.

The subgrid physics implemented in the hydro simulations is captured by five subgrid model parameters $\theta_\mathrm{sub} = \{\kappa_\mathrm{w}, e_\mathrm{w}, M_\mathrm{seed}, v_\mathrm{kin}, \epsilon_\mathrm{kin}\}$. The first two describe a galactic outflow model, with $\kappa_\mathrm{w}$ being the wind velocity and $e_\mathrm{w}$ being the energy outflow. The other three parameters describe the AGN model, with the black hole seed mass, $M_\mathrm{seed}$, and two kinetic feedback parameters $v_\mathrm{kin}$, $\epsilon_\mathrm{kin}$. The parameter ranges are provided in Table~\ref{tab:params1}. The sampling of the parameter values is based on a symmetric Latin hypercube design~\citep{mckay1979_lhs}. More details about the simulations and the parameter variations can be found in~\citet{ramachandra2026_emulator}.

\begin{table}[ht]
\caption{Ranges of the astrophysical subgrid model parameters $\theta_\mathrm{sub}$ for the hydrodynamic simulations. Parameters without units are dimensionless; $M_\mathrm{seed}$ is quoted in $h^{-1}\mathrm{M}_\odot$ as used internally by CRK-HACC.}

\begin{tabular}{llcc}
\toprule
\text{$\theta_\mathrm{{sub}}$} & & \text{min($\theta_\mathrm{{sub}}$)} & \text{max($\theta_\mathrm{{sub}}$)} \\ 
\midrule
$\kappa_\mathrm{w}$ & & 2 & 4 \\
$e_\mathrm{w}$ & & 0.2 & 1 \\
$M_\mathrm{seed}$ & $[h^{-1}\mathrm{M}_{\odot}]$ & $0.6 \times 10^6$ & $1.2 \times 10^6$ \\
$v_\mathrm{kin}$ & $[\mathrm{km\,s^{-1}}]$ & $0.1 \times 10^4$ & $1.2 \times 10^4$ \\
$\epsilon_\mathrm{kin}$ & & $0.2$ & $12$ \\ 
\bottomrule
\end{tabular}

\label{tab:params1}
\end{table}

\subsection{Data Products for Gravity-Only Simulations}
\label{sec:products}

We provide a range of data products from both snapshots at redshifts $z = \{0, 0.1, 0.5, 1.0, 2.0\}$ for Frontier-E GO, the Discovery simulations, and the SciDAC suite; Last Journey snapshot coverage is
described in Section~\ref{sec:lastjourney}. For Frontier-E GO, we additionally provide lightcone catalogs covering $0 \leq z \lesssim3$, as well as lightcone maps (Section~\ref{sec:lc_maps}). The information includes halo catalogs, halo profiles, and particle data for high-mass halos. The combination of products enables a wide variety of scientific work, and we provide a Python library to work with these products cohesively without requiring manual data management. Information about the halo and galaxy catalogs, particle-level data, and sky maps is provided below.

\subsubsection{Halo Catalogs}
\label{sec:halocatalogs}

The data release includes complete halo catalogs for all simulations for the redshift snapshots listed above. Halos in HACC are first identified using a friends-of-friends (FoF) algorithm~\citep{davis_1985} with a linking length of $b=0.168$. Following this, halos above a certain mass limit are again measured using a spherical overdensity (SOD) algorithm. A sphere is centered on the potential minimum of the FoF halo and increased in size until its average density drops below a threshold value of 200 times the critical density of the Universe. We measure a large range of halo properties for both the SOD and FoF particle sets, including the mass, position, velocity, kinetic energy, velocity dispersion, circular velocity, angular momentum, and shape measurements. For SOD halos we additionally provide the radius and the halo concentration determined from profile fitting. Halo catalogs include both the FoF and SOD halo measurements.

\subsubsection{Halo Profiles}
\label{sec:haloprofiles}

Halo profile measurements are provided for SOD halos. These give spherically averaged properties of all particles within a number of radial bins, starting at the minimum potential point and extending out to twice the halo radius as measured by the SOD algorithm discussed in Section~\ref{sec:halocatalogs}. These property bins include information such as particle counts and radial velocities. 

The OpenCosmo toolkit automatically associates halos stored in the halo catalog with their profiles. This makes it easy to express complex analysis tasks that require both the high-level properties available in the catalogs and one or more of the measured profiles.

\subsubsection{Halo Particles}
\label{sec:haloparticles}

Access to the full particle information associated with high-mass halos is provided, typically restricted to large objects with masses $\sim 10^{14}\,\mathrm{M}_{\odot}$ and above. The scale of the particle data makes it impractical to store all particles in the simulation, but this cutoff allows in-depth studies of cluster-mass objects. These datasets include positions, velocities, and particle ID tags. 

As with profiles, the OpenCosmo toolkit automatically links particles to the halo's high-level properties available in the halo catalog. Users may iterate over the halos and their particles one by one, enabling them to analyze halos at scale without overwhelming their system's available memory by loading excessive particle data.

\subsubsection{Lightcone Catalogs}
\label{sec:lc_cat}

Lightcone (observer-frame) catalogs detail halos at the redshift and location at which an observer placed at the simulation origin would detect them, allowing for more direct comparisons to cosmological surveys. For the gravity-only Frontier-E simulation we apply a halo finding algorithm identical to that discussed in Section~\ref{sec:halocatalogs}, with the exception that it is performed on particles in the lightcone frame. The measured FoF and SOD halos are identical in both threshold sizes and property lists to those in the snapshot frame, with the addition of angular coordinates, redshift, and an index describing the lightcone box replication. As for the snapshot frame, profile and high-mass halo particle information is provided alongside the halo catalogs.

In simulations run prior to 2025, such as Last Journey, halo properties were computed solely for the snapshot frame as described in Section~\ref{sec:halocatalogs}, and then moved to the lightcone using interpolation through the halo merger trees, as described in Section~4.3 of~\citet{2021ApJS..252...19H}. 

\subsubsection{Lightcone Maps}
\label{sec:lc_maps}

Map-level data of lightcone particles can be used to enable studies of the matter field in an observer frame, as well as to derive gravitationally induced weak lensing deflections of photons. We take lightcone particles which are computed in situ during the simulation evolution, bin them into fine redshift slices, and project them onto a high-resolution HEALPix map~\citep{gorski_2005} using a cloud-in-cell (CIC) kernel~\citep{1988csup.book.....H}. For release, the maps are then summed into shells of width $\sim$145\,Mpc to more easily allow for access and use in ray-tracing algorithms. For the Frontier-E simulation, these maps are computed using all available particles up to a redshift of $z=3$, and then using a 1\% subsample in the range $3<z<5$, capturing the vast majority of nonlinear effects.

\subsection{Data Products for Hydrodynamic Simulations}
All quantities that are available in gravity-only halo catalogs and profiles are also available in hydrodynamic catalogs with the addition of information regarding the distribution and properties of gas and stellar populations. A number of hydrodynamic-specific quantities (e.g., metallicity, temperature, mass) are provided for individual tracer particle species as relevant.

\subsubsection{Halo Catalogs, Halo Profiles, and Halo Particles}

Halo particles in the hydrodynamic simulations contain the same position and velocity information as particles in the gravity-only simulations, but grouped by species. The available types of particles are dark matter, gas, star, and AGN. As mentioned above, individual particle species have additional properties, such as age for stars, internal energy for gas, and accretion rate for AGN.

Note that FoF halo finding algorithms are not well defined in the context of multiple particle types, so the FoF halo finder in the case of hydrodynamic simulations is run on the dark matter particles only. In practice this introduces two subtleties: 1) the FoF mass contains only the dark matter mass and must be scaled by the baryonic fraction to obtain an approximate mass equivalent to the gravity-only simulations, and 2) to obtain hydro-relevant quantities one must use SOD halos.

\subsubsection{Galaxy Catalogs}
\label{sec:galaxy_catalogs}

Galaxies are first identified in CRK-HACC simulations using the DBSCAN clustering algorithm~\citep{1996kddm.conf..226E} applied to star particles. DBSCAN is a density-based algorithm that generalizes FoF finders by incorporating a minimum number of neighbor particles. After identifying galaxy locations in this way, we then measure properties at this position over all particles within a fixed aperture radius in proper units, typically $\sim50$\,kpc. For more specifics on galaxy finding see Section~2.9 of~\citet{frontiere2025_galaxyformation}.

The galaxy catalogs include basic kinematic measurements, masses of the major components, and additional properties such as helium fraction and metallicity. We retain the aperture-defined star particle set for detailed galaxy inspections. All of the data products described in this section are accessible through the OpenCosmo portal and its clients, which we describe next.

\section{The OpenCosmo Portal and Clients}\label{sec:tools}

The OpenCosmo web portal combines a set of high-level query tasks for requesting custom subsets of data from the underlying simulations with a user interface designed to guide users to the data most relevant to their science. In this section, we describe the design philosophy behind the portal's user experience, the query tasks currently available, the portal's browser-based visualization capabilities, the programmatic interfaces built on the same infrastructure, and the current access requirements.

\subsection{Design Philosophy and User Experience}\label{sec:ui}

The portal frontend is designed to mirror the philosophy of the project as a whole: scientists should be able to interact with their data in terms of high-level, familiar science concepts, rather than writing custom query code. Data portals often provide an SQL-like interface to the underlying data. While such query languages are powerful, most use cases involve relatively simple operations on a small subset of the available columns. The portal therefore pre-selects the quantities most relevant to scientific analysis -- halo masses and concentrations, for example -- so that common queries can be expressed directly in these terms.

\begin{figure*}[t]
\centering

\begin{subfigure}{\textwidth}
    \centering
    \includegraphics[width=0.6\linewidth]{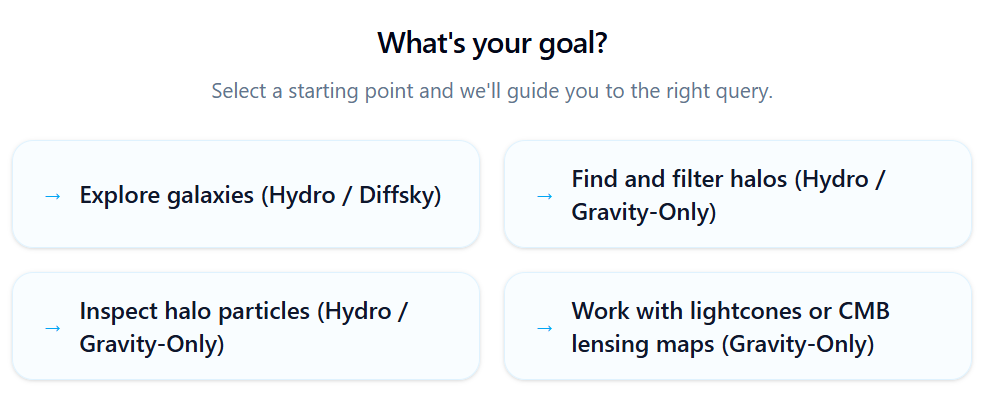}
\end{subfigure}\hfill
\vspace{1em}
\begin{subfigure}{\textwidth}
    \centering
    \includegraphics[width=0.6\linewidth]{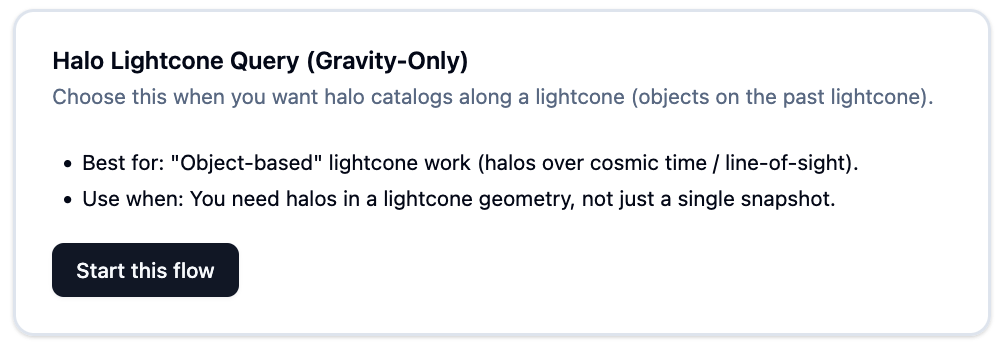}
\end{subfigure}

\caption{Top: High-level goal buttons available to the user on the OpenCosmo web portal's homepage. Bottom: One of the query cards displayed after clicking on the ``Work with lightcones or CMB lensing maps (Gravity-Only)'' goal.}
\label{fig:frontend}
\end{figure*}

The frontend is built around guided discovery. The homepage organizes the available capabilities by science goal; selecting a goal (e.g., ``Explore Galaxies (Hydro/Diffsky)'') leads to a page that summarizes the relevant datasets and query tasks -- their purpose, parameter guidelines, and typical use cases -- from which the user can launch the corresponding query directly (Figure~\ref{fig:frontend}). Technical terminology is defined inline through hover-activated tooltips backed by a glossary of technical terms, so that users unfamiliar with simulation-specific vocabulary do not need to leave their workflow to consult external references.

The portal also retains a complete record of each user's runs, including the query task, target dataset, input parameters, and per-stage execution status. Any previous run can be reopened and resubmitted with modified parameters. Beyond convenience, this record provides lightweight provenance -- the exact inputs that produced a given output file remain available -- and makes systematic parameter variations straightforward.

Ultimately, \emph{how} we provide access to data is as important as which data
we provide access to. The technical complexity hidden behind this interface is
substantial; we describe how it is managed in Section~\ref{sec:technical}.

\subsection{Querying Capabilities} 

Data access on the portal is organized into query tasks. Each task exposes a set of parameters appropriate to the data being retrieved, such as mass or redshift ranges, along with optional extensions; for example, including the particle data associated with each returned halo. All tasks are built on top of a common data management framework, which makes new query workflows straightforward to create and deploy. The queries currently available on the portal are:

\begin{itemize}[itemsep=1pt, topsep=2pt, leftmargin=*]
\item \textbf{Halo Query:} Queries the simulation's halo catalog based on quantities such as mass, concentration, and radius. For the hydrodynamic simulations, users may choose to include the galaxies associated with each halo. Particles and measured profiles are available for higher-mass halos. The OpenCosmo toolkit automatically links these data types together so that users may address the halo as a single object in their analysis.
\item \textbf{Galaxy Query:} For hydrodynamic simulations, queries the galaxy catalog based on cuts in quantities such as stellar mass and gas mass. Users may choose to also return the host halos associated with the galaxies, in which case all sibling galaxies within the host halo will be included in the output.
\item \textbf{Particle Query:} The Particle Query allows querying particles based on specific halo IDs. A user may choose to retrieve a large halo catalog using the Halo Query, identify objects relevant to their work based on high-level properties, then query the particles for the halos of interest for further study.
\item \textbf{Map Query:} Queries maps of synthetic observables built on HEALPix grids, such as integrated matter density or weak lensing convergence.
\item \textbf{Synthetic Galaxy Query:} Queries synthetic galaxy catalogs produced using the Diffsky pipeline as described in Section~\ref{sec:diffsky}. Unlike the galaxies directly provided from the CRK-HACC simulations, the Diffsky catalogs are produced with a model that is undergoing continuous improvement and will be updated accordingly.
\item \textbf{Analysis Plots:} Several tasks are available for producing publication-quality plots backed by the datasets described above, with built-in comparisons to other datasets. Plots are generated with the HAvoCC (HACC Analysis and Validation to Observational Constraints Code) code\footnote{\url{https://git.cels.anl.gov/hacc/HAvoCC}}.
\end{itemize}

Data retrieved from the portal (with the exception of analysis plots) are returned as HDF5 files formatted to be used with the OpenCosmo toolkit. The toolkit can perform additional cuts on the data, compute new quantities, associate related datasets such as halos and their particles, and perform spatial queries, among other common analysis tasks. Crucially, the portal queries themselves are performed with the toolkit on raw data which are stored in the same format as the data that are returned to the user. This allows us to develop new queries rapidly using a common API, and ensures that the output of any query can be read and analyzed with the same tooling.

\subsection{Visualization}\label{sec:viz}

\begin{figure*}
    \centering
    \includegraphics[width=\textwidth]{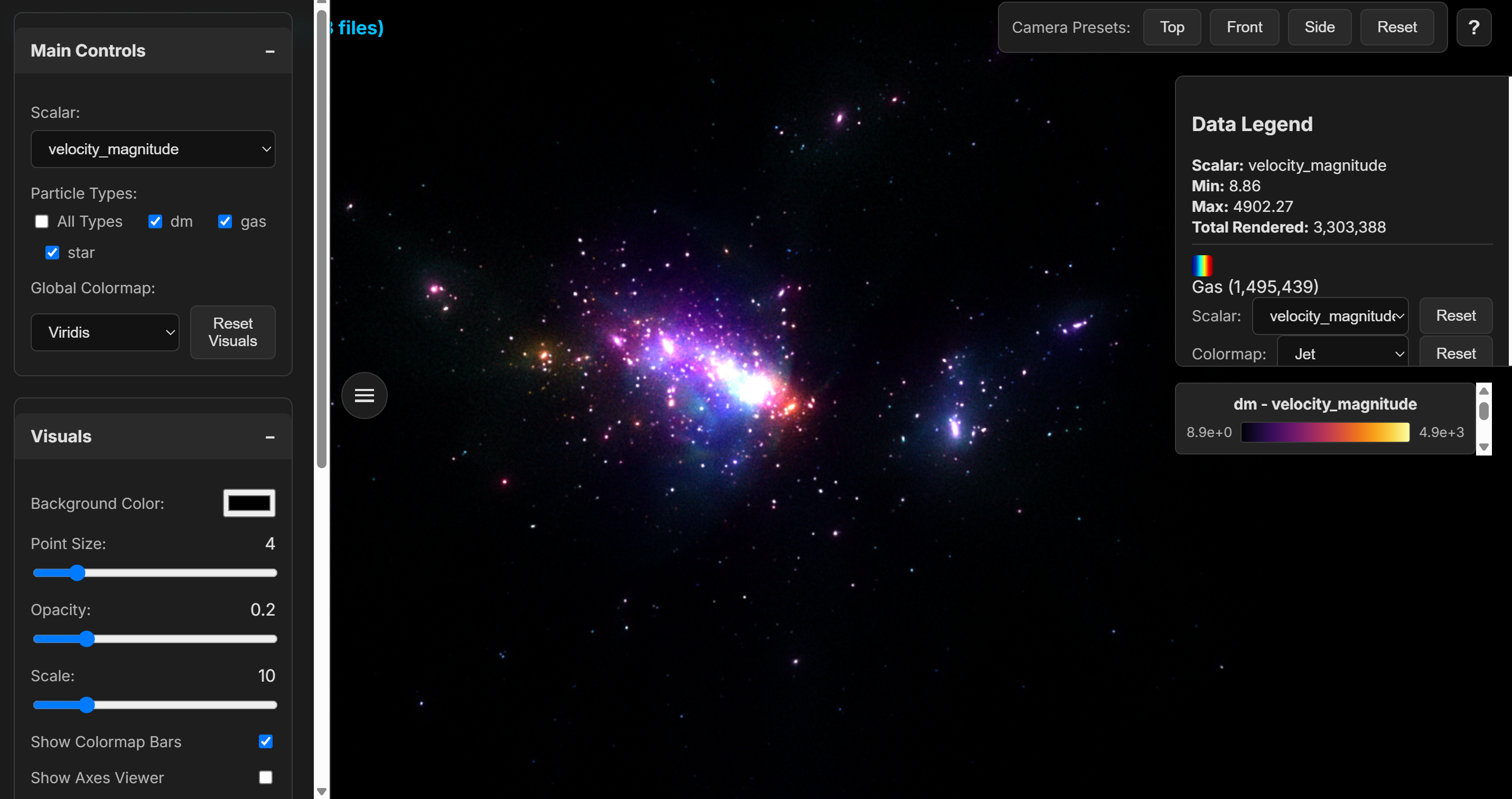}
    \caption{Screenshot of the web-based 3D viewer displaying dark matter, stars, and gas. Individual particles are colored by the magnitude of their velocity.}
    \label{fig:overview} 
\end{figure*}

The portal includes an interactive 3D viewer that allows users to explore individual halos directly, as shown in Figure~\ref{fig:overview}. The viewer runs entirely in the browser; no dedicated desktop application is required.

A standalone query task retrieves data for visualization; on completion, the data are converted to Parquet\footnote{\url{https://parquet.apache.org}} format and ingested into the viewer automatically. The viewer supports the following operations on the particle data:

\begin{description}[style=unboxed,itemsep=1pt,topsep=2pt, leftmargin=*]
    \item[Coloring] Particles are colored by mass by default, or by any measured property such as velocity magnitude or temperature.
    \item[Filtering] Particles may be filtered on properties, for example to isolate the hot gas in cluster-scale halos.
    \item[Slicing] Particles may be hidden on one side of a user-defined spatial slice.
    \item[Selections and statistics] Particles within a user-defined box may be selected dynamically, with summary statistics generated automatically.
    \item[Brushing and linking] Selecting a range of bins in a per-type scalar histogram highlights the corresponding particles in the 3D view in real time, linking structural features to specific scalar value ranges.
    \item[Immersive VR]
    Users with a WebXR-compatible headset may enter an immersive VR session directly from the browser, with the headset view mirrored to the desktop display so that collaborators can follow the exploration; an in-headset control panel exposes the most frequently used display parameters.
\end{description}

The camera system provides preset orientations, panning, and user-selected pivot points for centering on and rotating about specific structures of interest.

\subsection{Programmatic Access: Command-Line Client and Agent Integration}
\label{sec:programmatic}

All capabilities of the web portal are also available programmatically. The \texttt{ocp} command-line client authenticates through the same Globus-based flow as the portal and allows users to browse the available datasets and query tasks, submit queries, monitor running jobs, and retrieve results directly from a terminal or shell script.

The platform additionally supports AI-agent-driven workflows via the Model Context Protocol (MCP)\footnote{\url{https://modelcontextprotocol.io}}. A remote MCP server exposes each query task as an individually callable tool, with input schemas generated on a per-user basis from the same dataset registry that drives the web portal (Section~\ref{sec:registry}); a local server with the same capabilities can be launched from the command-line client. Because the served schemas are standard JSON Schema, host applications can validate agent-issued requests before submission, and the server only advertises datasets the authenticated user is authorized to access. These interfaces have been used to build and test agentic frameworks that perform scientific analyses semi-autonomously.

\subsection{Accounts and Authentication}\label{sec:access}

Access to OpenCosmo is currently limited to users with an account at NERSC or the ALCF, or an affiliation with a DOE National Laboratory. Because the portal triggers jobs on machines at these facilities, we cannot currently offer fully public access. Users may request access by filling out a form linked on the portal homepage. All applications are subject to review to ensure they meet facility requirements.

Authentication is provided through Globus Auth, and authorization is managed through Globus group membership: when a user logs in, their group memberships determine which datasets and query tasks are available to them. Users who are not yet members of an approved group are prompted to request access when they first load the portal; requests are reviewed against the eligibility requirements above and granted by adding the user to the appropriate group. Basic information about users, such as their name and institutional affiliation, is collected for security and compliance purposes.

Group membership also allows us to differentiate access beyond this baseline. For example, a dataset can be made available to a selected set of users before its public release; once the data are public, a single configuration change makes it accessible to all users of the portal. The same mechanism can accommodate facility-specific access requirements as new sites are added.

Authentication requirements are, however, likely to change rapidly in the coming year. In particular, the American Science Cloud (AmSC) aims to provide federated identity across many DOE facilities. This authentication model would enable us to expand access, and we plan to support this authentication method when it becomes available. We will continue to seek ways to expand the availability of the platform in compliance with facility requirements.

\section{Technical Discussion}
\label{sec:technical}

The OpenCosmo portal is built on several overlapping technologies, which combine into a coherent and extensible data analysis framework. The goal of this framework is to build a \textit{domain-specific platform} on top of \textit{domain-agnostic patterns}. Separating (wherever possible) the parts of the platform that must have knowledge of astrophysics and cosmology from the parts that do not ensures much of the framework can be easily transferred to new science domains while ensuring the user-facing layer feels familiar to astrophysicists and cosmologists. 

\subsection{Overview}

The OpenCosmo portal provides an interface for querying subsets of large simulation datasets in a completely automated way. The web frontend is written in TypeScript using the Svelte framework\footnote{\url{https://svelte.dev/}}. The frontend software itself does not hard-code information about the queries it is capable of performing or the datasets it is working with, ensuring it can be easily modified to support other science use cases. Queries are defined by swappable JSON-formatted \emph{task definitions}, which specify the expected inputs and all relevant user-facing text (e.g., titles and descriptions). Task definitions do not embed any information about specific datasets. Instead, they carry small binding annotations (e.g., ``this field selects a dataset that provides halo catalogs'' or ``this field's options are supplied by the selected dataset''), which the portal resolves at request time against a dataset registry (Section~\ref{sec:registry}). This separation allows easy addition, removal, or modification of both queries and datasets, or even the creation of a new portal backed by an entirely different set of data. 

Submitting a query on the portal triggers a job on one of the various compute resources where we host data. This delegation is transparent to the user, who only needs to specify which dataset they are interested in querying. The orchestration layer operates in terms of datasets, facilities, and jobs, not cosmology. The framework delegates work based on the dataset requested by the user, but simply passes on any domain-specific parameters to the data tooling in the following step. The query jobs themselves are submitted to a job scheduler in the usual way, but their small sizes, combined with allocation policies, typically result in turnaround times of minutes rather than hours. 

The query itself is performed by the OpenCosmo toolkit: a domain-specific query tool that understands astrophysics/cosmology data, in particular the relationships between different data types.  Once the query is complete, the result is returned to the user as an HDF5 file, which can be read and further queried with the same toolkit (Section~\ref{sec:toolkit}). The toolkit environment is containerized, ensuring it can easily be deployed on new resources.

We now discuss the individual pieces of the framework in detail. 

\subsection{The Dataset Registry}
\label{sec:registry}

Information about the available datasets is maintained in a single, authoritative registry that lives alongside the query and orchestration tooling. Each dataset is described by one structured metadata file, containing its name, a short description and usage guidance, physical facts such as box size and particle masses, the list of available snapshots and lightcone redshift ranges, access requirements, and a set of \emph{capability tags} declaring which data products exist on disk (e.g., halo catalogs, halo particles, HEALPix maps).

Task definitions declare their data requirements as expressions over capability tags, and the portal computes at request time, for each user, which datasets a given task can operate on, which optional inputs are available for the selected dataset, and the valid values and ranges of every dataset-dependent parameter. Datasets a user is not authorized to access remain visible but locked, with a hint describing how access can be requested (Section~\ref{sec:access}).

Crucially, the portal treats the registry contents as opaque data: all astrophysics- and cosmology-specific semantics are confined to the metadata files and the query tooling. This is a concrete example of the separation between domain-specific and domain-agnostic components discussed above, and the same registry machinery could serve an entirely different scientific domain without modification.

\subsection{OpenCosmo Data Format and Toolkit}
\label{sec:toolkit}

The foundation of all query tasks is raw simulation data and our query software. The OpenCosmo Data Format is a specification for storing astrophysics and cosmology data in HDF5 files. It provides support for units, cosmological parameters, spatial indices, metadata, and linking of related data types such as halo properties and halo particles. Although the format is presently limited to simulated datasets, support for observational data will be made available in the future.

Support for accessing, querying, and performing analysis tasks on data stored in this format is provided by the OpenCosmo toolkit, which automatically handles file metadata and interfaces with standard tools like Astropy~\citep{astropy_2022} and NumPy~\citep{numpy}.  It has built-in support for MPI that is transparent to the user, and can leverage parallel HDF5 for rapid data output. A script that runs serially on a laptop can be run across multiple nodes of an HPC system without changes to the script's logic or control flow. The system is designed to be approachable for early-career users who are still learning data tooling, but capable of supporting data-intensive workflows in a massively parallel environment.

\begin{figure*}[t]
\begin{minted}{python}
    import opencosmo as oc
    import numpy as np
    import astropy.units as u
    
    ds = oc.open("haloproperties.hdf5", "haloprofiles.hdf5")

    def nfw(c, r_200, r_norm, rho_crit):
        md = 4/3 * np.pi * rho_crit * 200 * r_200**3
        Anfw = np.log(1 + c) - c / (1 + c)
        dmdr = md * r_norm / (r_200 * Anfw) / (1 / c + r_norm) ** 2
        return dmdr
        
    def get_nfw_fit(halo_properties, halo_profile, cosmology, redshift):
        rho_crit = cosmology.critical_density(redshift).to(u.Msun / u.Mpc**3).value
        r_200 = halo_properties["sod_halo_radius"]
        r_norm = halo_profile["sod_halo_bin_radius"] / r_200
        best_fit = nfw(halo_properties["sod_halo_cdelta"], r_200, r_norm, rho_crit)
        return {"best_fit_nfw": best_fit, "residual": best_fit - halo_profile["sod_halo_bin_mdelta"]}
        
    ds = ds.evaluate(get_nfw_fit, 
        halo_properties = ["sod_halo_radius", "sod_halo_cdelta"], 
        halo_profile = ["sod_halo_bin_radius", "sod_halo_bin_mdelta"],
        cosmology = ds.cosmology,
        redshift = ds.redshift,
        format = 'numpy', 
        insert = True
    )

\end{minted}    
\caption{Example code for computing the best-fit NFW profile and residuals from the measured profile with the OpenCosmo toolkit. The \texttt{nfw} function computes an ideal NFW profile based on the halo concentration, the radius $r_{200c}$, the normalized halo radius bins, and the critical density of the Universe. The \texttt{get\_nfw\_fit} function computes these quantities from the stored data, passes them to the \texttt{nfw} function, and returns the results. The \texttt{ds.evaluate} call provides the information necessary for the toolkit to retrieve the relevant data and manage the computation.}
\label{fig:toolkit_code}
\end{figure*}

The toolkit is, at its core, a lazy-query engine: data are (when possible) not loaded into memory until actually needed. Users may perform filters, add and drop columns, and define analysis tasks while leaving most of the data on disk. When data are requested, the toolkit instantiates the dataset in memory, taking into account all transformations the user has performed.

In addition to basic query tasks, the toolkit supports various routines for generating new data from existing columns in a managed way. It supports column expressions, which are directly inspired by the Polars dataframe library\footnote{\url{https://pola.rs/}} (and recently adopted by Pandas\footnote{\url{https://pandas.pydata.org/}}). These expressions are unit-aware, and, like access to the raw data, fully lazy. 

The toolkit also supports the fully managed evaluation of complex expressions that cannot be expressed with column arithmetic. Figure~\ref{fig:toolkit_code} shows an example code snippet where a user computes an ideal Navarro--Frenk--White profile~\citep{Navarro_1997} based on the concentration and radius of a halo, then uses it to compute the residuals from the actual halo profile as given by HACC. The majority of the code in the snippet is cosmology-related. The toolkit manages the computation from start to finish, collects the results, and inserts them into the dataset for the user to access (and later save to disk, if they choose). The user only needs to specify which data are necessary for the computation to be performed, allowing the toolkit to minimize the amount of data read from disk. In an MPI environment, this computation is automatically distributed across all available ranks.

The scale-agnostic nature of the toolkit accelerates the ability to build the platform as well as the users' ability to perform their scientific work. At present, all queries available on the web portal are pre-specified by the OpenCosmo developers. The toolkit allows developers to specify arbitrary queries rapidly, while guaranteeing the output will be easily usable. In the future, we plan to allow users to submit their own analysis code to selected systems.

\begin{figure*}
    \centering
    \includegraphics[width=\textwidth]{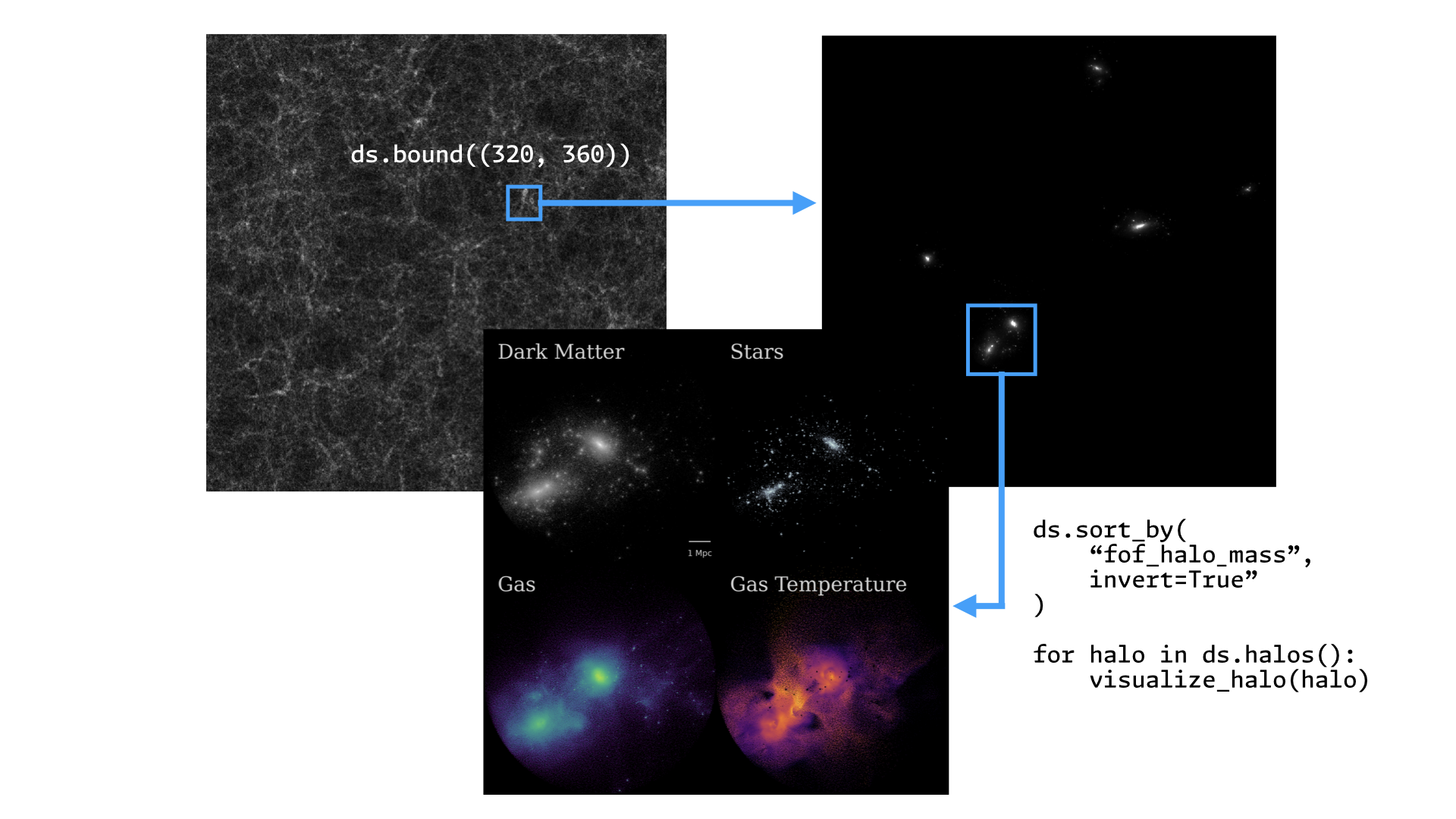}
    \caption{Demonstration of transition from a full simulation to a single structure of interest. Top left: Density field of a $512$\,Mpc subvolume of the Frontier-E hydrodynamic simulation. Top right: Cluster-mass halos within a $40$\,Mpc sub-region of the same volume. Bottom: Image generated from particle data of the most massive halo in the $40$\,Mpc subfield. The code shown is the actual code used to perform this task.}
    \label{fig:toolkit-scale} 
\end{figure*}

The toolkit also recognizes that cosmological work often relies on information at many different scales, and provides tools for easily moving between these scales as the analysis demands. Data are typically opened for an entire simulation at once, and are then filtered down based on user requests either in parameter space (e.g., by filtering on halo mass) or in physical space (by performing spatial queries). Once the user has a subset of the data they are interested in, they can quickly and easily extract relevant data, such as the particles for a given halo of interest. Because the toolkit evaluates queries lazily, this approach works well even in cases where the memory available to the system is only a small fraction of the total data volume. An example of this ``scale transitioning'' behavior is shown in Figure~\ref{fig:toolkit-scale}. This efficient approach allows us to run highly complex queries on our largest datasets using only a single compute node.

The OpenCosmo toolkit provides a unified base layer for describing queries and transformations on the underlying data. However, in order to build a user-facing platform on top of these capabilities, we need an orchestration layer for delegating and managing workloads, which we discuss in Section~\ref{sec:arch}.

\subsection{Multi-Facility Task Orchestration Framework}
\label{sec:arch}

\begin{figure*}
  \import{globus_illustration}{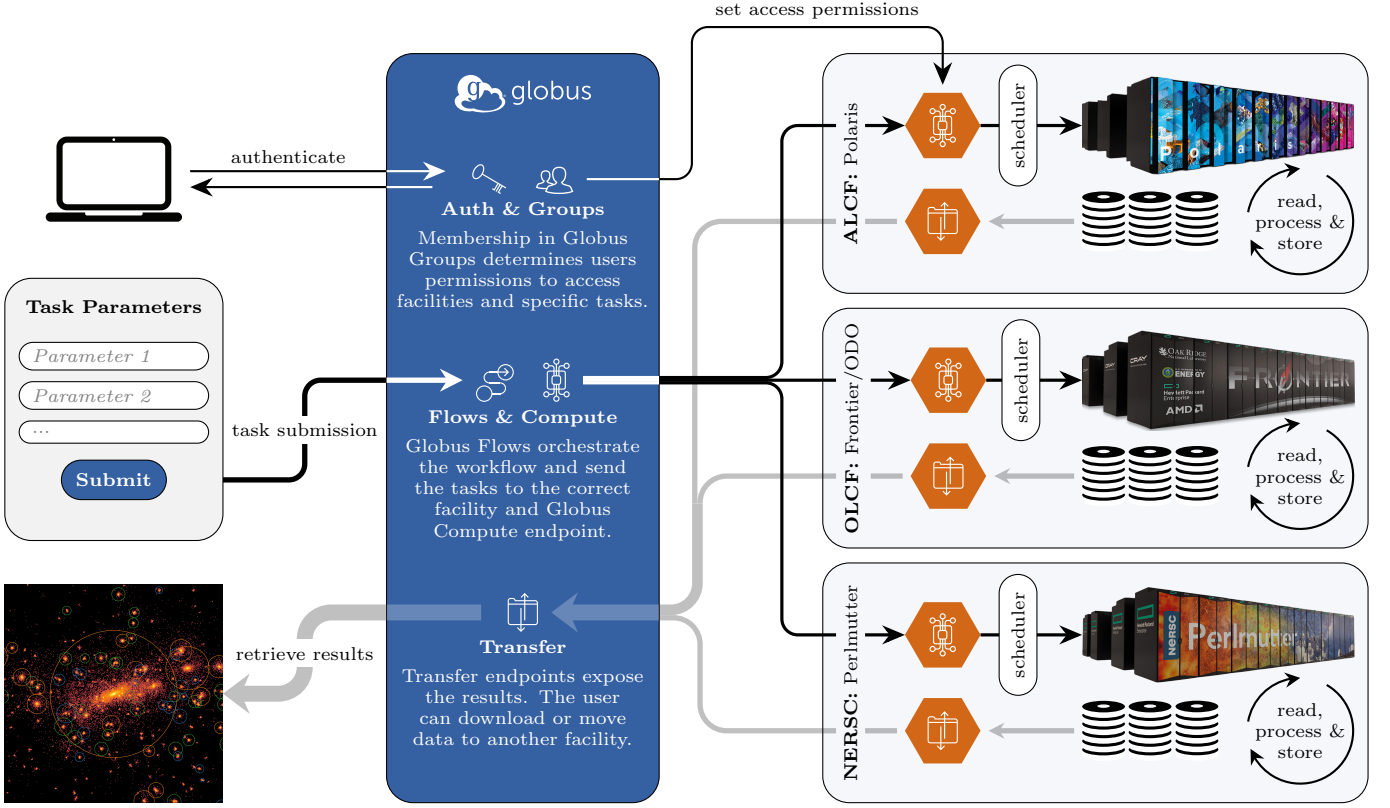}
  \caption{
    Illustration of the OpenCosmo workflow. On the OpenCosmo portal, users authenticate through Globus with an approved identity provider. Depending on membership in various OpenCosmo Globus Groups, users have access to different facilities and tasks. Task submission is routed through Globus Flows to Globus Compute endpoints running at the facilities, where the tasks are scheduled and executed and the resulting dataset is stored on a publicly accessible Globus Transfer collection. The results can be previewed on the OpenCosmo portal directly, downloaded through HTTP, and transferred through Globus to another facility.
  }\label{fig:opencosmo_diagram}
\end{figure*}   

The OpenCosmo platform is built on top of Globus, which provides capabilities for user authentication, workflow orchestration, and data transfer. This framework takes query tasks defined using the toolkit and abstracts them into generic workflows that can be handled with a single system. Individual tools available to the user can be added, removed, or modified without changing any aspect of the core architecture. Building user-facing services on top of traditional HPC resources has historically presented significant challenges, some of which can now be addressed via capabilities offered by Globus, as discussed below. Figure~\ref{fig:opencosmo_diagram} illustrates the Globus integration in the OpenCosmo framework.

\subsubsection{Lifecycle Management with Globus Flows}

When the user submits a query, the parameters for the query are collected and submitted to the Globus Flows platform~\citep{chard2023_globus_flows}. The platform manages the request until it either fails or returns data to the user. The user can track the progress of their job in the portal directly. Users can run multiple jobs at the same time, although we enforce per-user rate limits to prevent a single user from monopolizing the available compute resources.

The Globus Flows platform provides a framework for combining computing and data transfer tasks across multiple facilities into a single managed pipeline. Flows support branching and error checking, which allows the platform to delegate work to multiple compute facilities and to fail quickly if it encounters a problem. In the event any given step fails, the failure is communicated to the user and displayed in the portal. 

The Globus Flows platform provides facilities for branching and introspection of the results of previous steps, but access to arbitrary code must be delegated to an appropriate action provider. We make use of the Globus Compute platform~\citep{chard2020_funcx} to perform task delegation and to monitor jobs that are running on HPC systems, which we discuss in the next section.

\subsubsection{Compute Endpoints}

The core hardware abstraction in Globus Compute is the Globus Compute endpoint~\citep{chard2020_funcx}: a persistent service that executes work on the hardware where it runs, from sub-second function calls to jobs submitted to the scheduler and run across many nodes.
Endpoints can execute arbitrary code, but may also be restricted to pre-registered functions taking user-provided parameters, in which case they operate much like serverless function-as-a-service platforms offered by commercial cloud providers (e.g., AWS Lambda, Oracle Cloud Functions, or Azure Functions).

OpenCosmo users never interact with the endpoint directly. The portal submits query payloads to the appropriate endpoint on the user's behalf via a service account managed by the OpenCosmo team. We use Compute endpoints for three distinct purposes: two manage queries at runtime and are described in the remainder of this section, while the third performs maintenance tasks and is discussed together with our deployment infrastructure in Section~\ref{sec:cicd}.

\subsubsection{Endpoint Resolution}

As data live at multiple facilities, the first step of performing any query is identifying where the requested dataset lives and delegating work to the appropriate facility. This task does not interact with raw data and therefore can be run anywhere with access to the internet. We implement this delegation using a Globus Compute endpoint running on a virtual machine. This endpoint also performs checks of all user-provided data to ensure they are properly sanitized before reaching the HPC system housing the data.

The endpoint runs a single function that takes the parameters of the query as input and returns the private UUIDs of a compute endpoint and a pre-registered query function at the facility hosting the requested dataset, together with a signed payload. This information is returned to the Globus Flow instance, which delegates the work accordingly; the facility endpoint independently verifies the signature and re-validates all parameters before execution. If the endpoint determines the parameters are invalid, it ends the flow immediately and returns an error code to the user. 

\subsubsection{Query Evaluation}

Once the system has determined where the query should be evaluated, the flow submits the query to an endpoint running on the HPC system that houses the data. The query payload includes the name of the simulation, the type of query to be performed, and any user-specified parameters. The endpoint automatically submits a job to the local scheduler, which calls the appropriate functionality and passes the user-provided parameters. 

All queries are performed using the OpenCosmo toolkit in a containerized environment. The endpoint receives access to a set of configuration files that describe how to run the underlying query workflow, allowing the code that runs on the endpoint itself to be generic. This architecture ensures we can easily deploy new workloads within the system with very few changes to the infrastructure layer. 

Our containerized environment is built on top of a generic image with Python, \texttt{mpi4py}~\citep{dalcin2021mpi4py}, and \texttt{h5py} with parallel HDF5 support. These container images are publicly available\footnote{\url{https://github.com/AstroPatty/parallel-hdf5}}, and are known to work with the runtime and MPI environments at the National Energy Research Scientific Computing Center (NERSC) and the Argonne Leadership Computing Facility (ALCF).

\subsection{CI/CD and Maintenance Tasks}\label{sec:cicd}

New features, improvements, and bugfixes are tested and deployed automatically through a Continuous Integration and Continuous Deployment (CI/CD) pipeline. Applying this standard practice to HPC resources at multiple facilities is challenging, as no mature CI frameworks target these environments.

Containers for query workflows are built on a virtual machine and pushed to Docker~Hub. Once this process completes, a maintenance endpoint at each facility housing our data is triggered to pull the new container(s) and prepare them for use. Preparation depends on the facility: NERSC's \texttt{podman-hpc} can run Docker images directly, while ALCF uses Apptainer as its primary container runtime, which cannot run Docker containers directly and must first convert them to the appropriate format. We have written a unified container management layer for our maintenance endpoints. The compute functions that run on the endpoint delegate to our library, which determines the necessary steps based on configuration passed as environment variables. This ensures a single function is deployed to all our maintenance endpoints, simplifying the overall deployment process. 

\subsection{Visualization Internals}

The interactive 3D viewer introduced in Section~\ref{sec:viz} combines three key strategies to load and process large datasets: an efficient columnar storage format, off-main-thread parsing, and on-the-fly computation of derived quantities.

The viewer fetches and loads simulation data outputs stored in the Apache Parquet format, which is well-suited for large-scale data because of its efficient compression and fast access to specific columns such as spatial coordinates. The renderer remains responsive while visualizing millions of particles. Particle types are detected at load time, so each species can be colored and filtered independently.  

To maintain responsiveness, data loading and parsing occur in a separate Web Worker, running independently from the main browser thread. Inside the worker, a WebAssembly (WASM) module, \texttt{parquet-wasm}\footnote{\url{https://github.com/kylebarron/parquet-wasm}}, reads and converts Parquet data into an Apache Arrow\footnote{\url{https://arrow.apache.org}} table, enabling fast in-memory access to the dataset. Once parsing is complete, all typed arrays -- including particle positions, type indices, and scalar buffers -- are transferred to the main thread via the browser's Transferable Objects mechanism. This zero-copy transfer avoids structured-clone overhead, allowing the full dataset to reach the rendering pipeline without duplicating memory. 

Once the data are loaded, the worker derives additional scalar values -- such as \texttt{magnitude} computed from velocity components ($v_x, v_y, v_z$) -- to augment the dataset while preserving the integrity of the original files.

The visualization engine is powered by Babylon.js\footnote{\url{https://babylonjs.com}}, a WebGL-based rendering framework. The system uses a shader-based point cloud renderer that represents millions of particles as a single mesh with custom vertex attributes. The vertex shader performs spatial transformations, color mapping, and slicing operations, while the fragment shader applies Gaussian-style falloff functions to produce anti-aliased circular points. By treating the entire point cloud as a single GPU-managed object, the system minimizes draw-call overhead for large datasets.

\section{Conclusions}
\label{sec:conclusion}

The OpenCosmo portal and toolkit provide at-scale access to and analysis of flagship cosmological simulation data. The portal exposes complex data products through query tasks phrased in familiar scientific terms, allowing users to select the specific data subsets relevant to their investigation. The toolkit is a query and analysis framework that handles the complexities of astrophysical and cosmological data while maintaining best-practice data management. The same tooling supports multi-step pipelines running on full-scale simulation data across many compute nodes and small workflows running on a laptop; this ability to cross scales in data size and complexity transparently is a key feature of OpenCosmo.

The initial data release provides subsets of flagship HACC simulations on demand: snapshot and lightcone data products -- halo catalogs, profiles, and particles, HEALPix maps, and synthetic galaxy catalogs -- drawn from the Frontier-E gravity-only simulation, Last Journey, the Discovery simulations, and the SciDAC 128 SG5 hydrodynamic suite with its gravity-only companion (Table~\ref{tab:param}). Datasets provided through the portal (accessible at \url{https://opencosmo.science}) are in the OpenCosmo HDF5 format, ready for further analysis with the toolkit.

We are actively exploring additional features that we hope to make available to the community, including fully user-specified queries, the integration of observational data, and dynamic access to remote data in scripts and workflows through a Python API. This list is not exhaustive, and we welcome community feedback both on the current tools and on additional capabilities that would be useful. 
Several of these capabilities -- most notably user-submitted queries and workflows -- will require significant changes to our backend: serializing code, building environments on the fly, and managing substantially more metadata, as well as addressing the security concerns that running user-submitted code entails.

Independently of these plans, aspects of the technical stack may change meaningfully in the coming year. A federated authentication solution for multiple facilities is planned as part of AmSC (see Section~\ref{sec:access}), and several facilities are rolling out APIs for submitting and managing workloads on their machines, which would remove the need to maintain persistent endpoints there.

Although focused on astrophysics and cosmology data, the infrastructure layer of the OpenCosmo platform is generic and can be adapted to other science domains. Building user-facing services on top of HPC systems remains challenging; for example, the automated testing and deployment pipelines that such services require are not readily available there. This is not an inherent limitation of these resources; it reflects the need for reusable infrastructure components that accommodate both the constraints of HPC systems and the needs of application developers deploying on them. We plan to publicly release components of the OpenCosmo infrastructure stack in the near future.

The OpenCosmo portal and toolkit are production-ready -- the portal's backend, web frontend, and command-line client have all reached their 1.0 releases -- and we will continue to support and extend them in
collaboration with the community.

\begin{acknowledgments}
\textbf{Acknowledgments:}
Work at Argonne National Laboratory, NERSC, and ALCF was supported by the U.S. Department of Energy, Office of High Energy Physics. Argonne, a U.S. Department of Energy Office of Science Laboratory, is operated by UChicago Argonne, LLC under Contract No. DE-AC02-06CH11357.

This research used resources of the Argonne Leadership Computing Facility, which is a U.S. Department of Energy Office of Science User Facility operated under contract DE-AC02-06CH11357.

This research used resources of the National Energy Research Scientific Computing Center (NERSC), a Department of Energy User Facility (project hacc).

This research used resources of the Oak Ridge Leadership Computing Facility at the Oak Ridge National Laboratory, which is supported by the Advanced Scientific Computing Research programs in the Office of Science of the U.S. Department of Energy under Contract No. DE-AC05-00OR22725.

We gratefully acknowledge use of the Bebop cluster in the Laboratory Computing Resource Center at Argonne National Laboratory.

The authors record their debt to previous and current members of the  HACC team in generating the data sets served by the portal. For their contributions to this work, we gratefully acknowledge the efforts of JD Emberson and Nicholas Frontiere.

The authors wish to acknowledge extensive help and valuable discussions with members of the Globus team, in particular Ryan Chard and Kyle Chard, on the design of the Globus Flows and Globus Compute integration.

Generative AI was used in the preparation of this manuscript for light editing, reformatting, and debugging of LaTeX issues. 

\end{acknowledgments}

\bibliographystyle{aasjournal}
\bibliography{bibliography}



\end{document}